# Mechanical therapy as a potential "green" way to attack cancer disease

Li-Ting Yi [1] and Jing Liu [1, 2*]

[1] Department of Biomedical Engineering, School of Medicine,
Tsinghua University, Beijing 100084, China

[2] Beijing Key Lab of Cryogenic Biomedical Engineering,
Technical Institute of Physics and Chemistry,
Chinese Academy of Sciences, Beijing 100190, China

**\*Address for correspondence:**
Dr. Jing Liu
Department of Biomedical Engineering,
School of Medicine, Tsinghua University,
Beijing 100084, P. R. China
E-mail address: jliubme@tsinghua.edu.cn
Tel. +86-10-62794896
Fax: +86-10-82543767

## Abstract:

Mechanical force is tightly connected to human health status, and the occurrence of disease can generally be ascribed to certain loss of force balance. However, the role of mechanical approaches in tumor therapy is largely neglected, while the currently available cancer prevention and treatment methods are generally either expensive or just cause too much side effect. In this article, we present a systematic interpretation on a promising strategy which was termed here as "mechanical therapy" for the first time based on the fact that mechanical force closely accompanies the whole life (growth and death) of a cell, and plays a crucial role in biological functions. In order to mold the mechanical force as a practical tool for cancer therapy, we expound the effects of the mechanical force related to the tumors from molecule to tissue level and evaluate its feasibility for treatment purpose. It can be conceived that given enough investigations, the mechanical therapy may generate big potential to open new windows for cancer prevention, relieving or even destroying if well administrated. As an easy going and convenient method, a systematic and thorough study of mechanical therapy may clarify more options for future clinical practices. Besides, when patients encounter various forces in daily life, they can at least identify the good or bad forces for a better and healthy life style based on the mechanisms thus disclosed.

# 1  Introduction

With the increasing incidence, mortality and morbidity of cancer [1], scientists are striving to search various promising ways to fight the stubborn tumor. In fact, a basic understanding of tumor therapy is very much dependent upon surgical resection, chemotherapy and radiotherapy which are commonly used in clinics (Fig. 1). Although these widely adopted clinical therapies have contributed significantly to relieving the patients suffering and saving their lives, they are still far from being idealistic enough because of the serious side effects. So far, the eradication rate of a surgical resection is still rather low, and has been proven with high recurrence. Meanwhile, the chemotherapy and radiotherapy are usually going with low specificity, in the sense that the tumor as well as the surrounding normal tissues may be destroyed simultaneously which causes the unexpected damage of the constitution and immune system. Alternative methods such as thermal therapy and biological therapy are emerging, trying to avoid such side effects. Although these methods improve the treatment effect to some degree, there are still a series of technical challenges to overcome and the related complicated equipment and high cost generally make it hard to be widely available. The current treatment situation reminds people that tumor therapy still urgently needs to innovate. Among the many efforts ever made in clinical cancer researches, there lacks of a systematic and thorough discussion on the therapeutic role of mechanical force (Fig. 1). Bearing the important role of mechanical force in mind, this paper is dedicated to comprehensively draft a promising way to attack tumor which in a large extent can be termed as tumor mechanical therapy.

Recent advances on the relationship between tumor microenvironment and cancer disease have generated considerable attention and turned the efforts into understanding the mechanism and significance of microenvironmental factors. Compared to the tremendous former efforts on biochemistry, little is known about using mechanical force for tumor treatment. However, the interactions between solid stress of tumor microenvironment and the tumor growth attract great interest in some researchers. Bissell's lab has done much work to investigate the link between microenvironment and tumor [2~5]. Furthermore, they demonstrated that the correction of the microenvironmental cues might reverse the malignant phenotype [4]. In fact, over the whole life circle, cells are continuously exposed to various forces, and all the activities from molecule level to organs are under regulation of the mechanical force [6]. The changes of a tumor cell can be described as a journey of force development [7]. Force produced in the living cells over the growth and development process [8], in turn acts on cells [9]. It sets up a feedback loop due to the interaction between the force and cells [10]. There exists various force types in human cells such as hydrostatic pressure, shear stress, compression, traction and tension forces. Once pathological changes happen in the human body, the nature of these forces will change as well [11]. The physiological behaviors of cells are also influenced with the variation of the applied force. For exploring the relationship between solid stress and tumor, Helmlinger et al. [12] simulated tumor microenvironment using agarose matrices, and the tumor spheroid was inhibited inside at the stress of 45 to 120 mm Hg. Cheng et al. [13] further proved that high mechanical stress controlled the size of tumor spheroid through inhibiting the tumor cell proliferation and inducing apoptotic cells death. The relationship between mechanical force and tumor cells is not only embodied on their interactions mutually, but also on the physical property of the cell itself. And the stiffness of cells can be measured using atomic force microscopy. It has been proven that different tumor types demonstrate the common stiffness. Nevertheless, cancer cells and normal cells are with different mechanical stiffness [14, 15]. Moreover, the higher the invasive potential of cell, the less stiff it is [16].

Clearly, compared to the magnitude of the microenvironmental force, its impact to life mechanics is much greater. Then a question can be naturally raised that how to apply the external force and how much its effect will be? Starting from this basic concern, here we would propose the mechanical therapy with generalized purpose for tackling tumor based on mechanical force. And some basic mechanisms as well as practical strategies will be comprehensively interpreted. The principles of mechanical therapy are different from existing application of mechanics. For example, the way of HIFU (high intensity focused ultrasound) removing the tumor through focused ultrasound is violent [17], but mechanical force takes gentle and slow ways to destroy tumor. Nevertheless, one can take full advantage of the stronger anti-tumor response induced by mechanical HIFU. In addition, there has been some research disclosed that for the women diagnosed with breast cancer who received massage therapy regularly, their quantity of natural killer cells and lymphocytes was increased [18], the result of which exactly supports the view of mechanical therapy. In contrast, some works also demonstrated that massage could promote metastasis in mice [19]. Similar to massage, the process by which the external force affects the tumor and body is not well understood. According to such comparison, a systemic and quantitative study is required to control the final output. One should clarify the limit of the "good" and "bad" effects of mechanical force so as to achieve determinate and accurate therapy.

## 2  Basic Features of Mechanical Therapy

Overall, mechanical force has great potential in treating tumor. The benefits of such therapy could offer can be classified as follows: (i) The diverse forms of mechanical force, such as contact force and penetrable vibration of ultrasound, provide multiple alternative options for treatment, which may well meet the requirements of fighting cancer from superficial region to inner part of human body. (ii) The therapy can be implemented in both local and systemic level. Some tumor patients who received surgical resection still died due to metastasis. Therefore, it is believed that cancer is a systemic disease, which is often hard to tackle with only local therapy. Such difficulty can be well addressed in part by mechanical force with different forms. (iii) Mechanical therapy is a green way for clinical use without any toxicity. Although radiotherapy and chemotherapy are effective when they are operated as systemic ones, their side effects would destroy tumor and normal cells together which induce poorer body constitution. In contrast, mechanical process is generally gentle and noninvasive. In a sense, the essence of the mechanical therapy is a way that helps people do sports passively. (iv) The enhancement of immunity by mechanical effects will be helpful for people to remove traces of disease. In addition to its potential use in directing the treatment, mechanical force may also hold the key to caring for prevention of cancer. Mechanical therapy intentionally changes the movement of body from taking to receiving, increasing the immunity, smoothing pathway and improving physical function. (v) The related devices of mechanical force are more easily to be realized, which has promising potential to become point of care treatment.

# 3   Biological Mechanisms of Mechanical Force

It is well known that proteins and genes are essential for controlling the physical function by regulating cell behaviors. Mechanical force is also thought to contribute to the life activities, cell fate and tissue development from microscopic to macroscopic levels as well as gene regulation and signal pathways. Without instruments, people can only recognize and feel the obvious physical activity, however, besides these movements, the inner compositions are working at high speed mode under the quiet surface, and all these movements are unable to be separated from the mechanical force. In order to prove the feasibility of mechanical therapy, we expound the role of mechanical force involved in many key biological processes at crucial steps (Fig. 2) which are discussed in detail as follows.

## 3.1 Interactions between mechanical force and biological molecules

Protein is contained in all human cells, which serves as the basic element to maintain normal physiological functions of life. The disorder in protein metabolism can induce function imbalance and disease. The structural changes of functional protein influence its corresponding function, and the force can achieve this change as a potent factor. Considerable work has been carried out to research the differences of proteins in cancer and normal cells. Enzyme is a kind of special protein which catalyzes biochemistry reactions in living organisms, and the conformational changes of enzyme can be regulated by force [20], thus mechanical force is able to affect biochemistry reactions indirectly. The mechanical behavior mediated by protein in organisms is defined by a cyclic process of mechanosensing, mechanotransduction and mechanoresponse [21]. Importantly, this process is of critical importance to the behaviors of cell growth, differentiation, shape changes and death. Force sensing involves the alteration of protein conformation, which generates the effects of opening ion channels, unfolding matrix proteins, cytoplasmic protein unfolding, alterations of enzyme kinetics and catch-bond formation if subject to force [21]. Transduction of mechanical signal is mediated by membrane protein. Moreover, the force can not only influence the activity of membrane protein itself but also the related signal pathways. Only the signal gets through the membrane protein to reach the nucleus, the cell will respond and convey information to make mechanoresponse. There are also some inner influences including the tension dynamic changes of cytoskeleton and motor activity. The force processes through a series of proteins to receive, deform, transmit and response. Besides, another intracellular mechanical force process should be noted is cell division. Chromosome segregation and mitotic spindle formation happen during this process which need precise force manipulation at suitable phase. Now, it has been proven that the external mechanical force regulates the metaphase progression to some extent. This regulation is controllable, and different conditions will produce two sharp contrastive results of anaphase onset time [11]. The molecules are the basis of realizing the function of life activities. The accumulation of the molecular effects influenced by mechanical force will further lead to the change of cell behaviors. It is certain that the mechanical force can affect the activity at molecular level, which consists of the foundation for mechanical force therapy. The studies reported have shed light on the interactions between internal mechanical force and biological molecules. But there are tremendous works need to do to understand clearly the mechanism of the effects that external force would act on the tumor molecules.

## 3.2 Interactions between mechanical force and cells

Cell as the basic unit of structures and life activities contains various proteins inside to implement functions. So, as the mechanical force can change the biological molecules, it inevitably leads to the variation of cell activities including tumorigenesis, invasion and metastasis. When the tumor proliferates surrounded by the ECM (extracellular matrix), their biophysical interaction is reciprocal. The balance is adjusted between the contractile stresses exerted by tumor cells and the elastic resistance of the ECM (Fig. 3). As the higher proliferation rate of tumor cells contrasts to the limited space, in order to absorb more nutrients, the tumor cells tend to transfer through the ECM barriers with the steps of invasion and metastasis. During the migration, tumor cells move forward depending on the contractile forces generated by themselves. On the other hand, cells respond to the encountered force by modifying their behavior and remodeling their microenvironment [6]. The whole process progresses reciprocally with continuous feedback of mechanical force. To achieve the goal of tumor treatment, a possible way is to design the scheme that increases the solid stress of ECM to improve the constraint stress and reduce the physical migration. Further, more mechanisms are awaiting to be clarified.

3.3 Interactions between mechanical force and tissues

The most intuitive mechanical behavior happens at tissue level. There are different grades of cell development during human ontogenesis. The tissues with specific types significantly depend on the different mechanical forces generating from cells. Cells within tissue adjust the change by altering their geometry under the action of mechanical force, and in the process of interaction between the formation of tissue and mechanical behavior. The corresponding structure-function relationships achieve the tissue development [9]. Remarkably, the disease progression, such as the development of cancer, is strongly influenced by mechanical properties [11]. Although it is over a century since the contribution of physical forces to cells and tissue in the embryo was realized, there still lacks of a systematic understanding of utilization of mechanical force to treat disease. On some level, the mechanical force acting on the tissue directs its three-dimensional (3D) shape [22]. Meanwhile the force alters the growth rate of the surrounding cells [10]. Due to the different molecules, cells and mechanics, when diseased cells proliferate to the degree to form the tumor tissue, the matrix of mammary tumor often becomes stiffer than in normal tissue [23]. Therefore, whether the change of the stiffness would influence the progression of tumor is worthy of investigation. And to clarify this doubt, a feasible way is to intentionally apply the mechanical force to induce and test the resulted changes in the target tissues.

# 4  Forms of Mechanical Therapy

4.1 Direct mechanical force treatment

The effects of mechanical force on biology provide insight into its clinical application. However, the specific ways for this therapy to act efficiently need to be well designed. For direct mechanical force therapy, its implementation is more suitable for the superficial tumors (e.g., melanoma (Fig. 4)) and easy-operating tumors (e.g., breast cancer). Current research on mechanical force is more concerned with the influences of solid stress on tumor cells in microenvironments by using the in vitro models. The studies are investigated in both 2-D and 3-D models, while the 3-D ones better reflect the tumors in reality. Currently, tumor spheroids forming in agarose gels are popular 3-D test models, and some results regarding the relationship between the solid stress and tumor cells have been demonstrated which however are not sufficient enough [8, 12, 13]. To confirm the specific effects of treatments on human body, the experimental objects are more inclined to be animals, and the models in vitro just provide the range of parameters for in vivo research. The external mechanical forces in previous studies were changed by piston or the different concentration of gels [8, 12, 13, 24]. These methods are confined to the laboratory category, and the equipment is simple. There are few specialized devices to generate mechanical force, and the only commercial one is the FlexCell FX-5000 Compression system. And the high price ($34,780) makes it hard to apply widely [25]. With this fact, developing a new device to fulfill the need of controllable parameters with low cost will be necessary. For describing the mechanical force therapy vividly, a proper example is to regulate massage which is always used to ease fatigue. And it has been proven that massage does have effect on tumor [18, 19]. Although the outputs of some research are inconsistent, neither of them is wrong. The results depend on the form of massage, including the strength, frequency and the position. Inverse results remind us to quantify and control the mechanical force used in therapy, and different parameters should correspond to the explicit results. The directly applicable mechanical forces may include compression,

fluid force, friction, tensile force, and elastic force, etc. The intensity, orientation and frequency of force should be all considered. Research with respect to each force will raise a large amount of scientific issues.

## 4.2 Indirect mechanical therapy

Mechanical force therapy are not confined to the direct mechanical force alone. It also includes the indirect ones that generate mechanical force or influence the mechanical behavior of the cells or tissues. As an option, the electric, magnetic, acoustic, and thermal strategies have the potential to develop as indirect methods for tumor treatment, which will diversify and extend the area of mechanical therapy. Although tumor therapy utilizing the electric, magnetic, acoustic, and thermal energies are not new, we would put more focus here on their mechanical effect and propose some new annotation in future development.

Cellular mechanics is one of the key factors affecting the cell homeostasis. Different cells have different electrical properties [26]. When the cell emerges in an electric field, the local distortion will happen in cell and its surrounding region [27], and cells will change their size dynamically [28]. A same external electrical stimulation in different types of cells will bring varied influences on the mechanical parameters of cells. A 2 V/cm direct current electric field can change the cell elasticity, cell cytoskeleton and membrane mechanics [29]. Some research manipulates cells by electric methods using electrophoresis and dielectrophoresis which correspond to the changes of the charge and polarization of cells respectively [30]. Biological membranes are the gates for mass transport, and there are multiple free ions existing on both sides of them. The dynamically stabilized ionic gradients maintain the electrolyte homeostasis. On account of the free ions, the electric force can play a role with them. The oscillating force exerted by high frequency external electric field causes a forced vibration of free ions and influences the biochemical balance of the membrane and even the whole cell function further [31]. Since the electric force is able to affect the cells, so is the magnetic force. The similar results are obtained by using oscillating magnetic fields exerted on the free ions, as mentioned in electric field ones [32]. In order to make the magnetic method selective, auxiliary materials are added into the cells, such as the magnetic nanoparticles and magnetic fluids, which would help the magnetic field specifically target the tumor cells rather than the normal cells.

Ultrasound is a common technology as applied in diagnostic and therapeutic areas. Physical effect is one of its important roles, and a famous clinical application is ultrasound lithotripsy. High-intensity ultrasound destroys the tumor, much like surgery. And yet we have new hopes of ultrasound method by using its mechanical effects as a palliative therapy. In this respect, some investigations have demonstrated that low-intensity ultrasound is more inclined to kill the tumor cells compared with normal cells [33]. The benefits of ultrasound are its penetrating and controllable scope of therapy. There are so many mechanical processes from molecular to tissue level and ultrasound as an example producing mechanical force can certainly alter these processes and disturb or correct the cell behavior. The optimal intensity of ultrasound will achieve the desired effect.

Thermal methods also own the ability to contribute to mechanical force therapy, during which, some mechanical changes can be induced or accompanied to alter the cell function. It does not simply alter the carbohydrate and lipid metabolism, but even induce the change in gene expression and biochemical adaptive responses by heat stress [34, 35].

## 5  Conclusions and Future Perspective

Human body is an integrated organic system that every part works together and influences each other. For a healthy person, each part of the body is in relatively different states of forces, but in a balance to maintain homeostasis. If any problem appears on certain point, it may bring evident harm to the whole body. The important clues have been found from some existing works, which establish the convincing foundation for mechanical therapy of tumor. A systematic research on the relationship between mechanical force and tumor is still in infancy, especially applying the force as a therapeutic tool needs further exploration in the near future.

People generally face various forces in daily life, and may subject to large or small force unconsciously. For healthy people, it is not necessary to worry about non-invasive force. However, the situation may be quite different for those in bad physical status. Overall, mechanical force has dual character that different force may accelerate the exacerbation, but may also alleviate or even cure the disease. Accordingly, it is of great significance to figure out the laws of mechanical therapy so as to avoid the disadvantages and catch the opportunities in the hope of guarding human health. More research is still needed to distinguish the property of force in the sense to find its bad or good effect. A number of experiments should be carried out to explore the development of tumor and normal cells, tissue and body under different types and magnitudes of force in the coming future. Before the experiments in vivo, the numerical simulation is also necessary, and some parameters should be clarified, such as the magnitude of invasive force of tissue.

It is a long exploring course that various difficulties will be met. One of the important issues is that people of different age and sex have different compositions of tissue, which influences the effective value of force respectively. So the therapy cannot be implemented with invariable standard. Indeed, administrating with personalized treatment is worthy to be considered. Moreover, the following are a series of corresponding therapeutic devices and instruments needing to be developed. Under controlled force conditions by virtue of the special instruments, the therapeutic regimen provided can be modified for different people. We believe that once the powerful functions of mechanical therapy were disclosed in the coming time, more practices can be enabled. A systematic and thorough study will bring about novel therapeutic strategy of cancer for future clinics. The mechanical force knowledge for tumor fighting can help patients recognize the "bad" or "good" force and cope with specific mechanical effect for better daily life. The mechanical therapy is expected to be a powerful weapon against cancer in the near future.

## Acknowledgments:

This work is partially supported by the National Natural Science Foundation of China under Grant 51376102.

**Figures and legends:**

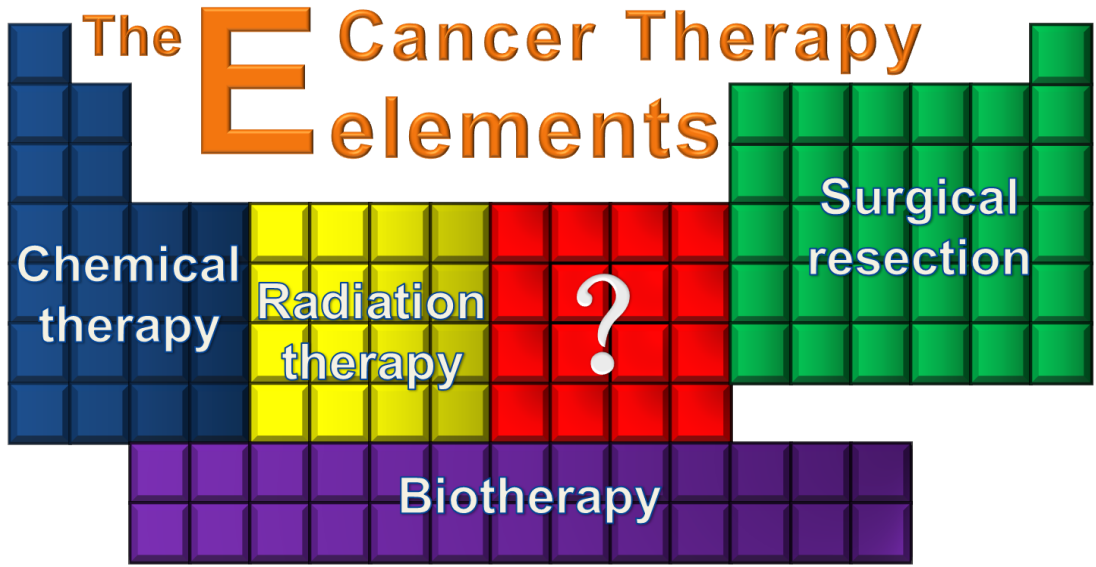

**Fig. 1** The elements of cancer therapy

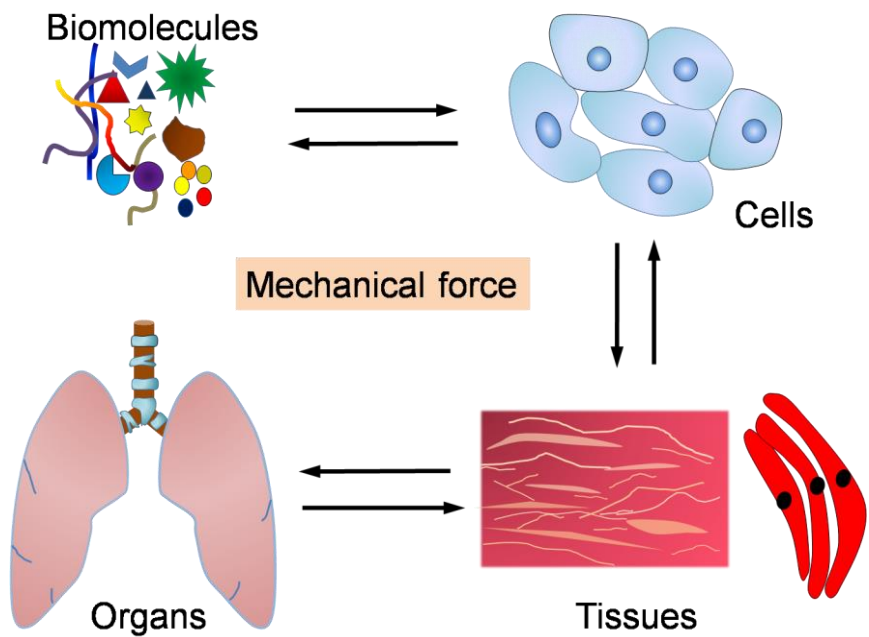

**Fig. 2** The interaction of mechanical forces among biomolecules, cells, tissues and organs

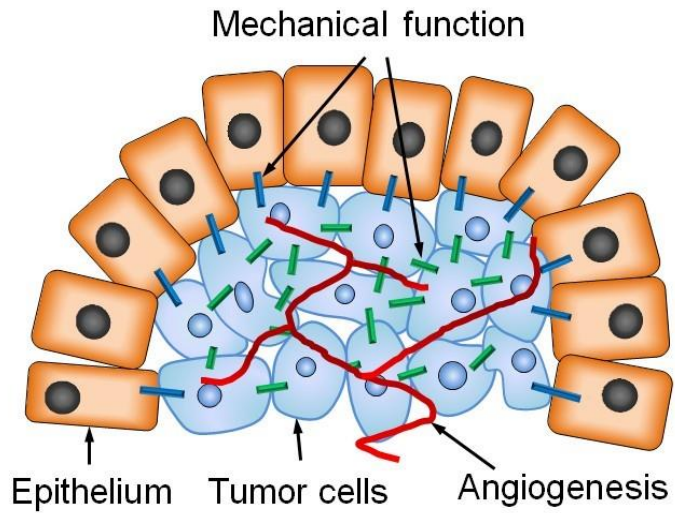

**Fig. 3** Compression of epithelium on tumor cells and the force between tumor cells themselves

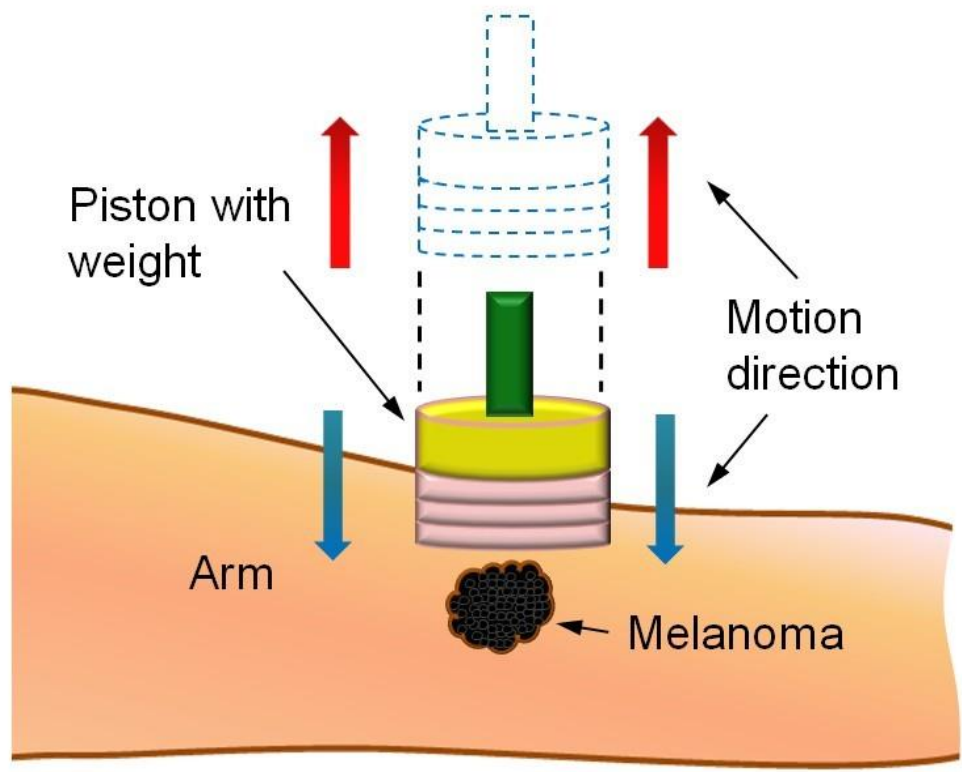

**Fig. 4** Schematic of mechanical force on melanoma